# CHANNEL ATTENTION RESIDUAL U-NET FOR RETINAL VESSEL SEGMENTATION

*Changlu Guo[1,*], Márton Szemenyei[1], Yangtao Hu[2], Wenle Wang[3], Wei Zhou[4], Yugen Yi[3,*]*

[1]Budapest University of Technology and Economics, Budapest, Hungary
[2]Hospital of the Peoples Liberation Army Joint Logistics Support Force, Nanchang, China
[3]Jiangxi Normal University, Nanchang, China
[4]Shenyang Institute of Computing Technology, Chinese Academy of Science, Shenyang, China
https://github.com/clguo/CAR-UNet

**ABSTRACT**

Retinal vessel segmentation is a vital step for the diagnosis of many early eye-related diseases. In this work, we propose a new deep learning model, namely Channel Attention Residual U-Net (CAR-UNet), to accurately segment retinal vascular and non-vascular pixels. In this model, we introduced a novel Modified Efficient Channel Attention (MECA) to enhance the discriminative ability of the network by considering the interdependence between feature maps. On the one hand, we apply MECA to the "skip connections" in the traditional U-shaped networks, instead of simply copying the feature maps of the contracting path to the corresponding expansive path. On the other hand, we propose a Channel Attention Double Residual Block (CADRB), which integrates MECA into a residual structure as a core structure to construct the proposed CAR-UNet. The results show that our proposed CAR-UNet has reached the state-of-the-art performance on three publicly available retinal vessel datasets: DRIVE, CHASE DB1 and STARE.

## 1. INTRODUCTION

Retinal vessel segmentation is of great significance in the early diagnosis of eye-related diseases. For example, Diabetic Retinopathy (DR) is a universal retinal disease caused by elevated blood sugar, accompanied by retinal vascular swelling [1]. However, the manual annotating of retinal vessels by ophthalmologists is a slow and labor-intensive task, so researchers have devoted themselves to proposing automatic retinal vessel segmentation methods.

*Corresponding author
This work was supported by the China Scholarship Council, the Stipendium Hungaricum Scholarship, the National Natural Science Foundation of China under Grants 62062040, and Chinese Postdoctoral Science Foundation 2019M661117.

In the past few decades, researchers have proposed a great number of methods for automatic retinal vessel segmentation, which are generally separated into two categories. One is image processing methods, which include pre-processing, segmentation, and post-processing steps, such as Bankhead et al. [2] proposed the use of wavelet transform method to enhance the detection of vessel foreground and background. The other is machine learning-based methods, which mainly use the extracted vector features to train a classifier to classify pixels in the retina. For instance, Lupascu et al. [3] designed 41-*D* feature vector for every pixel to train the AdaBoost classifier to classify each pixel in the retinal image.

Recently, deep learning-based methods have been used for automatic segmentation of retinal vessel and have achieved excellent results. Fu et al. [4] improved the vessel segmentation ability by using a convolutional neural network (CNN) with a Conditional Random Field (CRF) layer and a side output layer. Zhang et al. [5] introduced an edge-based mechanism in U-Net [6] to reach an improved performance. Wu et al. [7] introduced a Multi-Scale Network Followed Network (MS-NFN) for retinal blood vessel segmentation. Guo et al. proposed Dense Residual Network (DRNet) [8] to segment blood vessels in Scanning Laser Ophthalmoscopy (SLO) retinal images. In DRNet, a type of residual structure named Double Residual Block (DRB) is employed to deepen the network to obtain more complex semantic information, which has achieved a significant improvement. Although these deep learning-based methods have realized significant results, the interdependence between the feature channels was ignored. Later, Wang et al. [9] applied channel attention mechanism for "skip connections", and channel attention was also successfully applied in other medical image analysis tasks [10, 11]. However, these methods inevitably increase the complexity of the network due to dimensionality reduction operations when obtaining the attention map. And in [9-11] only average-pooling is used for aggregating spatial information, but max-pooling collects another important clue about distinctive object features to infer a finer channel-wise attention [12]. Therefore, in this work, we introduce a lightweight Modified Efficient Channel Attention that uses both average-pooled and max-pooled features simultaneously to further improve the performance of retinal vessel segmentation.

In this study, we propose a novel deep learning-based Channel Attention Residual U-Net (CAR-UNet) model, which achieves the state-of-the-art performance in retinal blood vessel segmentation of fundus images. Specifically, we have the following contributions: (1) Inspired by the recently proposed ECA-Net [13], which preserves performance while greatly reducing the network complexity in computer vision tasks (including image classification, object detection, and instance segmentation), we improve the ECA module in ECA-Net and name it Modified Efficient Channel Attention (MECA). (2) We apply MECA to "skip connections" for giving weight to each feature map from the contracting path, instead of copying them equally to the corresponding expansive path. (3) We consider the relationship between feature channels, so MECA is integrated on the basis of DRB [8] to propose Channel Attention Double Residual Block (CADRB). (4) Based on above work, we propose Channel Attention Residual U-Net (CAR-UNet) and evaluate it on DRIVE, CHASE DB1 and STARE datasets. The results demonstrate that our proposed CAR-UNet has reached the state-of-the-art performance on all three datasets.

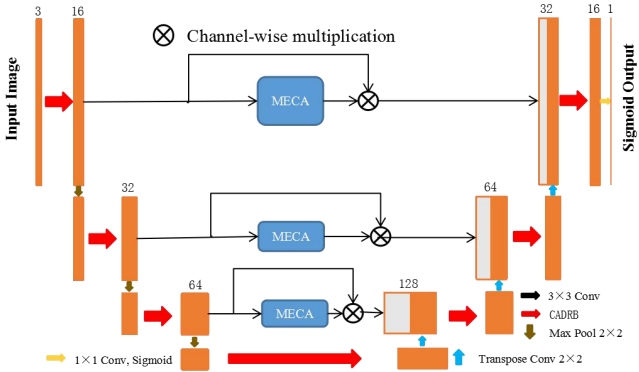

**Fig. 1**: The CAR-UNet architecture

## 2. PROPOSED METHODOLOGY

### 2.1. Network Architecture

The detailed architecture of Channel Attention Residual U-Net (CAR-UNet) is displayed in Fig. 1. The network structure of CAR-UNet is derived from U-Net, where the original convolution blocks are replaced by the CADRB. CAR-UNet contains two paths with the same number of residual blocks, namely the contracting path (left) and the expansive path (right). In the contracting path, each step includes a CADRB, and a 2×2 max pooling layer with step size of 2 is used for downsampling. Each step in the expansive path includes upsampling using transposed convolution, concatenating with the feature maps weighted by MECA from the contracting path, and then following a CADRB. In the last layer, we employ 1×1 convolution and Sigmoid activation function to get the required feature map.

### 2.2. Modified Efficient Channel Attention (MECA)

Channel Attention (CA) was first used as a *squeeze and excitation block* for classification [14], which generates channel attention maps by using the relationship between the channels. And recent works showed that the channel attention mechanism has great potential in improving the performance of deep convolutional neural networks (CNN). However, most methods dedicated to achieving better performance inevitably increase the complexity of the model. Wang et al. [13] proposed an Efficient Channel Attention (ECA) module that uses $1D$ convolution to avoid the dimensionality reduction operation in the *squeeze and excitation block*, which greatly reduces the complexity of the model while maintaining superior performance. However, in ECA only average-pooling is used for aggregating spatial information, but max-pooling collects another important clue about distinctive object features to infer a finer channel-wise attention [12]. Therefore, in order to aggregate the spatial information, we adopt both average pooling and maximum pooling to obtain finer channel-wise attention and named it Modified Efficient Chanel Attention (MECA) , as shown as Fig. 2. Formally, input feature $F \in R^{H \times W \times C}$ through the channel-wise max-pooling and average-poling can generate $F_{mp} \in R^{1 \times 1 \times C}$ and $F_{ap} \in R^{1 \times 1 \times C}$, respectively, e.g., at the $c$-th channel:

$$F_{mp}^c = Max \ (F^c(i,j)), 0 < c < C, 0 < i < H, 0 < j < W \quad (1)$$

$$F_{ap}^c = \frac{1}{H \times W} \sum_{u=1}^{H} \sum_{j=1}^{W} F^c(i,j), \ 0 < c < C \quad (2)$$

where $Max(\cdot)$ obtains the maximum number, $P^c(\cdot)$ represents the pixel value at a specific position of the $c$-th channel, and $H$, $W$, and $C$ stand for the height, width, and the number of channels of the input feature $F$, respectively. The two descriptors are then forwarded to a shared weight $1D$ convolutional layer to generate a channel attention map $M^c \in R^{1 \times 1 \times C}$. Then, MECA applies the channel-wise addition to combine the output feature vectors obtained by the shared $1D$ convolution. In short, the channel attention map is calculated as:

$$M(F) = \sigma(Conv1D(F_{ap}) + Conv1D(F_{mp})) \quad (3)$$

where $Conv1D(\cdot)$ represents the $1D$ convolutional layer and $\sigma(\cdot)$ denotes the Sigmoid function. It is worth mentioning that the kernel of the $1D$ convolution used by MECA in this paper is set to 3 (k=3) , that is, only 3 parameters are added for each MECA used, which directly proves that MECA is a very lightweight module.

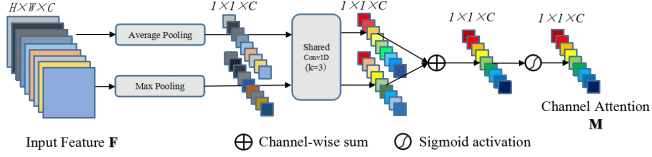

**Fig. 2**: Diagram of MECA

### 2.2. Channel Attention Double Residual Block (CADRB)

In order to extract more complex features in the retinal image, DRNet [8] proposed Double Residual Block (DRB) to build a deeper network. And for preventing network from overfitting, DRNet introduced DropBlock [15], a structured variant of dropout, which randomly discards the local block area and regularize the convolutional layer more effectively than dropout [15, 16]. Although DRNet only shows that it is an efficient method to segment blood vessels in Scanning Laser Ophthalmoscopy (SLO) retinal images, but we believe the structure of Double Residual Block is also applicable to fundus images.

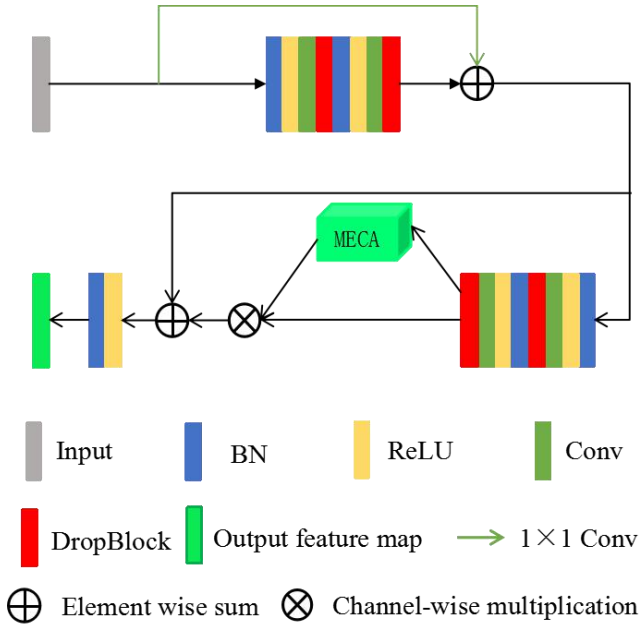

**Fig. 3**: Diagram of CADRB

Channel attention mechanism learns the importance of each feature channel through learning automatically, and uses the obtained importance to enhance features and suppress features that are not important to our retinal vessel segmentation task. In other words, MECA can extract channel statistics between channels, thereby further enhancing discriminative ability of the network.

Based on the above works and the recent success of *squeeze and excitation block* [14] and *convolutional block attention module* (CBAM) [12] in computer vision tasks, we integrate MECA into DRB and propose the Channel Attention Double Residual Block (CADRB), as shown in Figure 3.

## 3. EXPERIMENTS AND RESULTS

### 3.1. Datasets and Preparation

We employ three public retinal fundus image datasets to evaluate our model, they are DRIVE, CHASE DB1 and STARE. The DRIVE dataset is collected from a Dutch Diabetic Retinopathy (DR) screening project and contains 40 color fundus images with a resolution of 565×584 pixels. The dataset is partitioned into a training set of 20 images and a testing set of 20 images. CHASE DB1 comes from the British Children's Hearing and Health Research Project, which contains 28 fundus images with a resolution of 999×960 pixels, of which the first 20 images are utilized for training, and the last 8 images are used for testing. STARE (Structural Analysis of the Retina) contains 20 fundus images with a resolution of $700 \times 605$. However, there is no analogous division of training and testing sets made originally. Therefore, we adopt a 4-fold cross-validation method for training and testing stage. The manual annotations of both datasets provided by human experts can be utilized as the ground truth.

To fit our network model, we performed a simple processing on both datasets. We adjusted the size of DRIVE, CHAS DB1 and STARE to $592 \times 592$, $1008 \times 1008$ and $704 \times 704$ by padding it with zero in four margins, respectively. In order to acquire more reasonable results, we crop the segmentation results to the initial size when evaluating.

In order to enhance the robustness of the network, we adopt random rotation and horizontal, vertical and diagonal flips to augment the training images of DRIVE, CHASE DB1 and STARE datasets.

### 3.2. Network Configuration

As mentioned before, we partition the datasets into training and testing sets. In order to further monitor whether our model is overfitting, we randomly select 10% of augmented training images in all three datasets as the validation sets.

We train the proposed CAR-UNet from scratch utilizing the training set. For all datasets, we set the number of the feature channel after the first convolutional layer to 16 and utilize the Adam optimizer to optimize our network with binary cross entropy as the loss function. For DRIVE, the training batch size is 2, and a total of 100 epochs are trained with a learning rate of $1\times10^{-3}$. For CHASE DB1, the training batch is 1, and a total of 50 epochs are trained with a learning rate of $1\times10^{-3}$. For STARE, the training batch is 3, and a total of 80 epochs are trained with a learning rate of $1\times10^{-3}$.

For the setting of DropBlock, the size of the discard blocks is set to 7 and the dropout rates is set to 0.15 for all datasets.

**Table 1**: Results of Different Models on the **DRIVE**

| Metrics | Year | Spe | Sen | Acc | AUC |
|---|---|---|---|---|---|
| U-Net[6]* | 2015 | 0.9820 | 0.7537 | 0.9531 | 0.9755 |
| R2U-Net [17] | 2018 | 0.9813 | 0.7799 | 0.9556 | 0.9784 |
| MS-NFN[7] | 2018 | 0.9819 | 0.7844 | 0.9567 | 0.9807 |
| Yan et. al. [18] | 2018 | 0.9818 | 0.7653 | 0.9542 | 0.9752 |
| DEU-Net[9] | 2019 | 0.9816 | 0.7940 | 0.9567 | 0.9772 |
| DUNet [19] | 2019 | 0.9800 | 0.7963 | 0.9566 | 0.9802 |
| DDNet [20] | 2020 | 0.9788 | 0.8126 | 0.9594 | 0.9796 |
| **CAR-UNet** | **2020** | **0.9849** | **0.8135** | **0.9699** | **0.9852** |

Note: *the results are obtained from [17]

**Table 2**: Results of Different Models on the **CHASE DB1**

| Metrics | Year | Spe | Sen | Acc | AUC |
|---|---|---|---|---|---|
| U-Net[6]* | 2015 | 0.9701 | 0.8288 | 0.9578 | 0.9772 |
| R2U-Net [17] | 2018 | 0.9820 | 0.7756 | 0.9634 | 0.9815 |
| MS-NFN[7] | 2018 | **0.9847** | 0.7538 | 0.9637 | 0.9825 |
| Yan et. al. [18] | 2018 | 0.9809 | 0.7633 | 0.9610 | 0.9781 |
| DEU-Net[9] | 2019 | 0.9821 | 0.8074 | 0.9661 | 0.9812 |
| DUNet[19] | 2019 | 0.9752 | 0.8155 | 0.9610 | 0.9804 |
| DDNet [20] | 2020 | 0.9773 | 0.8268 | 0.9637 | 0.9812 |
| **CAR-UNet** | **2020** | 0.9839 | **0.8439** | **0.9751** | **0.9898** |

Note: *the results are obtained from [17]

**Table 3**: Results of Different Models on the **STARE**

| Metrics | Year | Spe | Sen | Acc | AUC |
|---|---|---|---|---|---|
| U-Net[6]* | 2015 | 0.9701 | 0.8288 | 0.9578 | 0.9772 |
| R2U-Net [17] | 2018 | 0.9820 | 0.7756 | 0.9634 | 0.9815 |
| Yan et. al. [18] | 2018 | 0.9846 | 0.7581 | 0.9612 | 0.9801 |
| DUNet [19] | 2019 | **0.9878** | 0.7595 | 0.9641 | 0.9832 |
| DDNet [20] | 2020 | 0.9769 | 0.8391 | 0.9685 | 0.9858 |
| **CAR-UNet** | **2020** | 0.9850 | **0.8445** | **0.9743** | **0.9911** |

Note: *the results are obtained from [17]

### 3.3 Segmentation results and evaluation

For the purpose that we can estimate the performance of our proposed CAR-UNet, the following metrics are employed: Specificity (*Spe*), Sensitivity (*Sen*), Accuracy (*Acc*), and Area Under the ROC Curve (*AUC*). The *AUC* can be used to measure the segmentation performance, because it is an indicator for evaluating the two classifiers and has the characteristic of not being affected by unbalanced data. If the value of *AUC* is 1, it means flawless segmentation.

In order to evaluate the performance of CAR-UNet, we compare it with several the exiting state-of-the-art models.

In Table 1, 2 and 3, we sum up the release year of each model and their performance on DRIVE, CHASE DB1 and STARE datasets. The results illustrate that, in three datasets, our CAR-UNet reaches the best performance among all competing models. Specifically, CAR-UNet has the highest *AUC* (0.45% / 0.73% / 0.53% higher than the second best), the highest accuracy (1.05% / 1.14% / 0.58% higher than the second best), the highest sensitivity, and specificity are comparable. The above results clearly demonstrate that CAR-UNet is a competent method for retinal vessel segmentation. Fig. 4 shows one sample result of the proposed CAR-UNet for DRIVE, CHASE DB1 and STARE images, respectively. From the figure, CAR-UNet can obtain clearer segmentation results even for very small blood vessels.

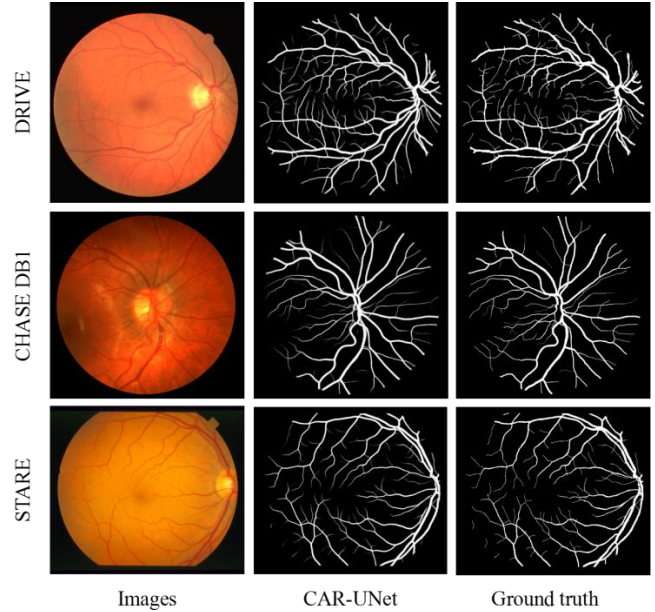

**Fig. 4**: Sample Segmentation results for three datasets

### 4. CONCLUSION

In this paper, we present a Channel Attention Residual U-Net (CAR-UNet) for retinal vessel segmentation of funds images. CAR-UNet considers the relationship between the feature channels, so a novel channel attention mechanism is introduced to strengthen the network's discriminative capability. Specifically, first, we introduce a Modified Efficient Chanel Attention (MECA) modify from the recently proposed Efficient Channel Attention (ECA). Then, we integrate MECA into Double Residual Block (DRB) to construct the contracting path and expansive path of the network. In addition, we apply MECA to "skip connections", assign weights to feature maps from the contracting path, instead of equally copying to the corresponding expansive path. Our experiments show that CAR-UNet reaches the state-of-the-art performance for retinal vessel segmentation on the DRIVE, CHASE DB1 and STARE datasets.